\documentclass[a4paper,10pt]{article}

\begin{document}
\title{\bf Inflation scenario via the
Standard Model Higgs boson and LHC\footnote{To the memory of John
Archibald Wheeler.}}
\date{}
\author{A. O. Barvinsky$^{1}$, A. Yu. Kamenshchik$^{2,\,3}$ and
A. A. Starobinsky$^{3}$}
\maketitle \hspace{-8mm} {\,\,$^{1}$\em Theory Department, Lebedev
Physics Institute, Leninsky Prospect 53, Moscow 119991, Russia\\
$^{2}$Dipartimento di Fisica and INFN, via Irnerio 46, 40126
Bologna, Italy\\
$^{3}$ L. D. Landau Institute for Theoretical Physics,
Kosygin str. 2, 119334 Moscow, Russia}

%\maketitle

\begin{abstract}
We consider a quantum corrected inflation scenario driven by a
generic GUT or Standard Model type particle model whose scalar
field playing the role of an inflaton has a strong non-minimal
coupling to gravity. We show that currently widely accepted bounds
on the Higgs mass falsify the suggestion of the paper
arXiv:0710.3755 (where the role of radiative corrections was
underestimated) that the Standard Model Higgs boson can serve as
the inflaton. However, if the Higgs mass could be raised to $\sim
230$ GeV, then the Standard Model could generate an inflationary
scenario with the spectral index of the primordial perturbation
spectrum $n_s\simeq 0.935$ (barely matching present observational
data) and the very low tensor-to-scalar perturbation ratio
$r\simeq 0.0006$.
\end{abstract}

\section{Introduction}
This is a challenging task to understand the nature of a
fundamental particle physics model that underlies the inflationary
scenario of the early Universe
\cite{S80,Guth,Sato,Linde,AlbSteinh,Lindechaotic} leading to
generation of scalar \cite{ChibMukh,H82,AlStar1,GP82,MFB} and
tensor \cite{S79} perturbations, the former leading to the
formation of observable structure of the Universe. Early attempts
to model inflation in terms of a self-interaction Higgs-like
scalar field $\varphi$ minimally coupled to gravity faced the
necessity to assume an extremely small coupling constant
$\lambda\sim 10^{-13}$ of its quartic self-interaction
$\lambda\varphi^4/4$ -- a natural candidate for the inflaton
potential motivated by particle phenomenology \cite{Lindechaotic}.
Initially this situation was considered very unfavorable from the
viewpoint of post-inflationary reheating, whereas now this simple
model is actually ruled out by the present observational data, see
e.g. \cite{WMAPforbidquarticinflation}.

It was also observed long ago that the problem of small $\lambda$
can be circumvented by adding to the Einstein term in the action
the non-minimal coupling term $\xi\varphi^2 R/2$ with a very large
coupling constant $\xi$, because in this case the CMB anisotropy
$\Delta T/T\sim 10^{-5}$ is given by the ratio $\sqrt\lambda/\xi$
\cite{Spokoiny,FM89,SalopekBondBardeen,FakirUnruh}. Therefore,
such smallness of $\Delta T/T$ can be obtained even for $\lambda$
close to unity (but still small enough to justify perturbative
expansion in $\lambda$) if $\xi$ is taken very large. This model
was considered from the viewpoint of quantum cosmology with the
tunneling cosmological wave function in \cite{qsi,BK,efeqmy},
where it was shown that quantum effects of matter fields are
crucial both for the formation of the initial conditions for
inflation \cite{qsi} and its dynamics \cite{BK}. Since inflation
may well be related to the GUT scale of particle physics, in
\cite{qsi,BK} the matter content of the model was taken to be of a
generic GUT type with the inflaton belonging to the scalar
multiplet of the GUT theory.

Recently it was advocated that in fact this non-minimally coupled
inflaton can be the Higgs boson of the Standard Model (SM), and no
new particles besides already present in the electroweak theory
are required to produce inflation with cosmological perturbations
in accordance with the CMB data \cite{BS}. This conclusion was
achieved within a tree-level approximation of this theory, because
its radiative corrections were claimed to be strongly suppressed
by a large value of the non-minimal coupling constant $\xi$.

The purpose of this paper is to show that this conclusion of
\cite{BS} is erroneous -- radiative corrections are actually
enhanced by a large $\xi$. They strongly affect the inflationary
dynamics of the Universe in a controllable way, and therefore can be
probed by current and future CMB observations and LHC experiments
testing SM. In particular, we will show that with a widely accepted
upper bound on the Higgs mass, $m_H\simeq 180$ GeV \cite{PDG}, the
model of \cite{BS} is falsified, but with $m_H\geq 230$ GeV the SM
Higgs can drive inflation scenario with a low spectral index
$n_s\geq 0.935$ and a very low tensor-to-scalar perturbation ratio
$r\simeq 0.0006$.

We consider the cosmological model with the classical Lagrangian
density
    \begin{eqnarray}
    &&{\mbox{\boldmath $L$}}(g_{\mu\nu},\varphi,\chi,A_\mu,\psi)=
    \left(\frac{m_{P}^{2}}{16\pi}
    +\frac{1}{2}\xi\varphi^{2}\right)R
    -\frac{1}{2}(\nabla\varphi)^{2}
    -\frac{\lambda}{4}(\varphi^2-\nu^2)^2\nonumber\\
    &&\qquad\qquad\qquad\qquad\quad
    -\frac 12 \sum_\chi (\nabla\chi)^2
    -\frac 14 \sum_A F_{\mu\nu}^2(A)
    -\sum_\psi \bar{\psi}\hat{\nabla}\psi\!
    \vphantom{\frac12}\nonumber\\
    &&\qquad\qquad\qquad\qquad\quad
    +{\mbox{\boldmath $L$}}_{\rm int}
    (\varphi,\chi,A_\mu,\psi)                       \label{model}
    \end{eqnarray}
containing the graviton-inflaton sector with a big non-minimal
coupling constant\footnote{For our choice of signs, the case of
the conformally invariant massless scalar field corresponds to the
coupling $\xi=-1/6$.} $\xi\gg 1$, and a generic GUT or SM sector
of Higgs $\chi$, vector gauge $A_\mu$ and spinor fields $\psi$
coupled to the inflaton $\varphi$ via the interaction term
    \begin{eqnarray}
    {\mbox{\boldmath $L$}}_{\rm int}
    =-\sum_{\chi}\frac{\lambda_{\chi}}2
    \chi^2\varphi^2
    -\sum_{A}\frac12 g_{A}^2A_{\mu}^2\varphi^2-
        \sum_{\psi}f_{\psi}\varphi\bar\psi\psi
    +{\rm derivative\,\,coupling},             \label{interaction}
    \end{eqnarray}
whose structure is dictated by the local gauge invariance. In
(\ref{model}) the inflaton $\varphi$ can be regarded generically
as a component of one of the scalar multiplets, which has a
non-vanishing expectation value in the cosmological quantum state.
After inflation it settles in the minimum of its potential at the
symmetry breaking scale $\varphi=\nu$. This scale is small enough,
$\nu^2\ll M_P^2/\xi$, so that at present the gravitational
interaction is mediated by the effective Planck mass squared
$M_{\rm eff}^2(\nu)=M_P^2+\xi\nu^2\simeq M_P^2$, $M_P\equiv
m_P/\sqrt{8\pi}=2.4\times 10^{18}$ GeV, and large enough to
generate the mass of the $\varphi$-particle,
$m_\varphi^2=2\lambda\nu^2$, which renders its interaction
short-ranged and not violating stringent bounds from Solar system
tests of gravity.

The paper is organized as follows. In Sec. 2 we present the quantum
effective action for the model (\ref{model}- \ref{interaction}),
previously obtained in \cite{BK}, and show that the contribution
from its radiative corrections was underestimated in \cite{BS}. In
Sec. 3 we investigate how the inflation dynamics of the model is
modified by these radiative corrections.  In Sec. 4 power spectra of
primordial scalar and tensor perturbations are presented and
compared with the recent WMAP bounds. It follows that the present
observational data put strong constraints on the parameters of the
model. In Sec. 5 we show that the Standard Model within the accepted
range of the Higgs mass does not satisfy these constraints, but it
can fit the CMB data for $m_H\geq 230$ GeV. We finish in Sec. 6 with
a general conclusion that the fate of the model may be resolved by
LHC tests of electroweak theory, and if nevertheless proved to be
viable, this model predicts a very small ratio of tensor to scalar
primordial perturbation $0.0006<r<0.001$.

\section{Effective action}
The quantum effective action for the model (\ref{model}) was
calculated in \cite{BK}. For a large and slowly varying mean
scalar field $\varphi$ and a large $\xi$, this calculation is
facilitated by the number of properties. Firstly, the non-minimal
coupling efficiently implies the replacement of the original
Planck mass parameter by $M_{\rm
eff}^2(\varphi)=M_P^2+\xi\varphi^2\gg M_P^2$. This means that the
contribution of the graviton and inflaton quantum loops is
essentially suppressed by powers of $1/M_{\rm eff}^2\sim
1/\xi\varphi^2$ \cite{BKK,BK}, and the main contribution comes
from quantum loops of matter sector of the model (\ref{model}) --
scalar fields $\chi$, vector bosons $A_\mu$ and spinor fields
$\psi$.

Secondly, due to the Higgs mechanism on the background of
$\varphi$, all these fields acquire large masses,
$m(\varphi)\sim\varphi$, following from the non-derivative part of
the interaction Lagrangian (\ref{interaction}):
    \begin{eqnarray}
    &&m_\chi^2=\lambda_\chi\,\varphi^2,\;\;
    m_A^2=g_A^2\,\varphi^2,\;\;
    m_\psi^2=f_\psi^2\,\varphi^2.               \label{masses}
    \end{eqnarray}
The scale of these masses are much larger than the characteristic
scale of the spacetime curvature $R\sim
\lambda(\varphi^2-\nu^2)^2/12M_{\rm eff}^2\sim
\lambda\varphi^2/\xi\ll\varphi^2$, and therefore the quantum
effective action can be found as a local $1/m^2$-expansion in
powers of the curvature, its gradients and the gradients of the
background scalar field $\nabla\varphi$. In the approximation
linear in $R(g_{\mu\nu})$ and $(\nabla\varphi)^2$, the answer
reads \cite{BK}
    \begin{equation}
    S[g_{\mu\nu},\varphi]=\int d^{4}x\,g^{1/2}
    \left(-V(\varphi)+U(\varphi)\,R(g_{\mu\nu})-
    \frac12\,G(\varphi)\,(\nabla\varphi)^2\right),         \label{effaction}
    \end{equation}
where the coefficient functions $V(\varphi)$, $U(\varphi)$ and
$G(\varphi)$ together with their classical parts contain one-loop
radiative corrections of the form
    \begin{eqnarray}
    &&V(\varphi)=\frac\lambda{4}(\varphi^2-\nu^2)^2+
    \frac{\lambda\varphi^4}{128\pi^2}
    \left(\mbox{\boldmath$A$}
    \ln\frac{\varphi^2}{\mu^2}+\mbox{\boldmath$B$}
    \right),             \label{effpot}\\
    &&U(\varphi)=
    \frac12(M_P^2+\xi\varphi^{2})+
    \frac{\varphi^2}{384\pi^2}\left(\mbox{\boldmath$C$}
    \ln\frac{\varphi^2}{\mu^2}+\mbox{\boldmath$D$}
    \right),                     \label{effPlanck}\\
    &&G(\varphi)=1+\frac{1}{192\pi^2}\left(\mbox{\boldmath$F$}
    \ln\frac{\varphi^2}{\mu^2}
    +\mbox{\boldmath$E$}\right).    \label{phirenorm}
    \end{eqnarray}

Here $\mbox{\boldmath$A$}$ is the following combination of Higgs,
vector gauge boson and Yukawa coupling constants of the GUT-inflaton
interaction Lagrangian (\ref{interaction})
    \begin{eqnarray}
    {\mbox{\boldmath $A$}} = \frac{2}{\lambda}
    \left(\sum_{\chi} \lambda_{\chi}^{2}
    + 3 \sum_{A} g_{A}^{4} - 4
    \sum_{\psi} f_{\psi}^{4}\right).   \label{A}
    \end{eqnarray}
\mbox{\boldmath$B$}, \mbox{\boldmath$C$}, \mbox{\boldmath$D$},
\mbox{\boldmath$E$} and \mbox{\boldmath$F$} are other five
combinations of the powers of these constants and their logarithms
\cite{Shore,AlStar1,BK}, whose specific form will be inessential
in what follows\footnote{In slightly different notations, these
combinations were obtained in \cite{BK} including the
non-logarithmic part of (\ref{phirenorm}). Logarithmic parts of
the coefficients \mbox{\boldmath$A$}, \mbox{\boldmath$C$} and
\mbox{\boldmath$F$} induced by one-loop radiative corrections from
vector fields were earlier presented in \cite{AlStar1} (see also
\cite{Shore}). }, and $\mu^2$ is a normalization point. On the
contrary, the constant (\ref{A}) is very important because, as we
will see, its effect is enhanced due to multiplication by $\xi\gg
1$. This constant determines the anomalous scaling behavior of the
theory, or its local conformal anomaly. In fact it arises as a
coefficient of $\varphi^4$ in the sum of quartic powers of
particle masses
    \begin{eqnarray}
    \frac1{64\pi^2}\,{\rm tr}\!\!
    \sum_{\rm particles} (\pm 1)\, m^4(\varphi)
    =\frac{\lambda\varphi^4}{128\pi^2}
    \mbox{\boldmath$A$},                               \label{Am}
    \end{eqnarray}
where summation takes into account boson/fermion statistics and the
trace is taken over spin-tensor indices.\footnote{With the masses
(\ref{masses}) this yields the expression (\ref{A}) in which the
coefficients 1, 3 and 4 imply respectively one, three and four
degrees of freedom of a real scalar field, massive vector field and
charged (Dirac) spinor field.} For $\xi\gg 1$ this quantity gives a
dominant contribution to the spacetime integrand of the one-loop
$\zeta$-function of the theory (subdominant contributions are
suppressed by powers of the curvature to mass squared ratio $\sim
1/\xi$) and also determines the coefficient of the logarithmic
Coleman-Weinberg potential in (\ref{effpot})
    \begin{eqnarray}
    {\rm tr}\sum_{\rm particles}
    (\pm 1)\,\frac{m^4(\varphi)}{64\pi^2}
    \,\ln\frac{m^2(\varphi)}{\mu^2}
    =\frac{\lambda\mbox{\boldmath$A$}}{128\pi^2}
    \,\varphi^4
    \ln\frac{\varphi^2}{\mu^2}+...\; .   \label{Jordanframe}
    \end{eqnarray}

Note that in the leading order of $\xi\gg 1$ the coefficient
$\mbox{\boldmath$A$}$ does not have a contribution from the
graviton-inflaton sector, which as was mentioned above is suppressed
by powers of $1/\xi\varphi^2$. In particular, the expression
(\ref{A}) does not contain in parentheses a typical contribution
$\lambda^2$ of the Higgs field itself.

Quantum corrections are obviously small for
$\mbox{\boldmath$A$}/32\pi^2\ll 1$, but the authors of \cite{BK}
claimed an additional mechanism of their strong suppression by a
large $\xi$, based on calculations in the Einstein frame of the
classical model (\ref{model}). This frame can be obtained by the
conformal transformation, $g_{\mu\nu}\to \hat g_{\mu\nu}$, $\hat
g_{\mu\nu}=\Omega^2(\varphi) g_{\mu\nu}$,
$\Omega^2(\varphi)=1+\xi\varphi^2/M_P^2$, and a relevant
reparametrization of the inflaton field, $\varphi\to\hat\varphi$,
rendering $\hat\varphi$ a canonical normalization of its kinetic
term. Under this transformation all the particle masses $m(\varphi)$
in the Einstein frame get rescaled as $\hat
m(\varphi)=m(\varphi)/\Omega(\varphi)\sim 1/\sqrt\xi$ and for
$\Omega\gg 1$ become very small and actually independent of
$\varphi$. This makes according to \cite{BS} their Coleman-Weinberg
potential small and very flat, so that quantum corrections are not
significant for the inflationary dynamics.

However, one should bear in mind that the contribution of this
potential to the effective action enters with the factor $\hat
g^{1/2}=\Omega^4(\varphi)\,g^{1/2}$ which cancels the dominant
effect of the decrease in mass,
    \begin{eqnarray}
    \hat g^{1/2}\,\hat m^4(\varphi)
    \ln\frac{\hat m^2(\varphi)}{\mu^2}
    =g^{1/2}\,m^4(\varphi)
    \ln\frac{\hat m^2(\varphi)}{\Omega^2(\varphi)\mu^2}
    \simeq {\rm const}\,g^{1/2}\,m^4(\varphi).
    \end{eqnarray}
Therefore, for large $\varphi$ quantum corrections calculated in the
Einstein frame differ from those of the Jordan frame
(\ref{Jordanframe}) by replacing the log factors with some
constants. This is not unusual that the covariant renormalization in
different conformally related frames leads to different results,
because the theory has a conformal anomaly, and this anomaly yields
a weak logarithmic frame dependence. This means that the mechanism
of suppression for radiative corrections advocated in \cite{BS}
should be much weaker -- instead of suppression by a power of the
conformal factor $\Omega^2\sim \xi\varphi^2/M_P^2$ the radiative
corrections in the Einstein frame are suppressed only by its
logarithm.

Below we will see that the dominant contribution of quantum
corrections to the inflaton rolling force originates from
differentiating namely the log factor in the effective potential
(\ref{effpot}). Therefore the disappearance of these factors in
the Einstein frame raises the question of which frame is
appropriate for calculating the quantum effects. The original
Jordan frame is the right one, because it determines the physical
distances in terms of the original metric $g_{\mu\nu}$. In
particular, physical (atomic) clocks measure the proper time of
just this frame. Covariant renormalization which introduces the
logarithmic factors and the normalization scale $\mu$ should be
performed in terms of this physical metric. This justifies the
original Jordan frame and the expressions (\ref{effpot}) and
(\ref{effPlanck}) containing the correct logarithmic terms.

\section{Inflation}
We apply the effective action (\ref{effaction})-(\ref{phirenorm})
with $\xi\gg 1$ to study inflationary dynamics in the range of the
inflaton field much beyond its current value at the minimum of the
classical potential, $\varphi^2\gg M_P^2/\xi\gg\nu^2$. Thus we
assume smallness of the following two parameters
    \begin{eqnarray}
    \frac{M_P^2}{\xi\,\varphi^2}\ll
    1,\,\,\,\,
    \frac{\mbox{\boldmath$A$}}{32\pi^2}\ll 1.     \label{smallness}
    \end{eqnarray}
It is also natural to assume that other combinations of coupling
constants are of the same order of magnitude as
$\mbox{\boldmath$A$}$, so that the second of bounds above also holds
for all of them
$(1/32\pi^2)(\mbox{\boldmath$B$},\mbox{\boldmath$C$},
\mbox{\boldmath$D$},\mbox{\boldmath$E$}, \mbox{\boldmath$F$})\ll 1$.

In fact (\ref{smallness}) guarantees the regime of the slow roll
approximation for the system (\ref{effaction}). In the leading order
corresponding to the omission of $\dot\varphi^2$ terms the equations
of motion read \cite{BK,KKT}
    \begin{eqnarray}
    &&\ddot{\varphi}+3H\dot{\varphi}-F = 0,             \label{KG1}\\
    &&H^2 = \frac{V}{6U} - \frac{U'F}{3U}.               \label{Hubble1}
    \end{eqnarray}
Here $H=\dot a/a$ is the Hubble parameter, $U'\equiv dU/d\varphi$,
and the rolling force $F=F(\varphi)$ in this approximation depends
only on $\varphi$ and reads\footnote{Strictly speaking, in the
leading order of the slow roll approximation $\ddot\varphi$ in
(\ref{KG1}) should be discarded on equal footing with
$\dot\varphi^2$, and it was retained only to emphasize a second
order derivative structure of the equation of motion. In
particular, the second term of (\ref{Hubble1}), which originates
from the $\dot a\dot U= aH\dot\varphi U'$ term of the Friedmann
equation in this model, results from expressing $\dot\varphi$ via
Eq.(\ref{KG1}) as $\dot\varphi=F/3H$.}
    \begin{equation}
    F(\varphi) = \frac{2 V U' - V'U}{GU + 3U'^2}. \label{force}
    \end{equation}
If the inequalities (\ref{smallness}) are satisfied, it reduces
to:
    \begin{eqnarray}
    F=-\frac{\lambda M_P^2}{6\,\xi^2}\,\varphi
    \left(1+
    \frac{\mbox{\boldmath$A$}}{64\pi^2}
    \frac{\xi\varphi^2}{M_P^2}\right)\equiv
    -\frac{\lambda M_P^2}{6\,\xi^2}\,\varphi
    \left(1+\frac{\varphi^2}{\varphi^2_I}\right),         \label{force1}
    \end{eqnarray}
where
    \begin{eqnarray}
    &&\varphi_I^2=\frac{64\pi^2 M_P^2}{\xi \mbox{\boldmath$A$}}.
    \end{eqnarray}
This expression for the rolling force was suggested in Eq.(6.7) of
\cite{BK} along with the scale of inflation field $\varphi_I$
which was derived from the principles of quantum cosmology --- the
probability distribution maximum at $\varphi_I$ of the quantum
corrected tunneling state of the Universe \cite{qsi}. Note that
while the second term in the right-hand side of (\ref{Hubble1}) is
small as compared with the first one, the second term in the
denominator in the right-hand side of (\ref{force}), just the
opposite, dominates the first one. Thus, during inflation the
effective Brans-Dicke parameter $\omega_{BD}\equiv U/2U'^2$ is
very small in this model.

For large $\varphi$ the strongest (cubic in $\varphi$) {\em
classical} term of the rolling force (\ref{force}) identically
cancels out and $F$ gets dominated by the second term of
(\ref{force1}). The latter is cubic in $\varphi$, too, but it is
essentially quantum, and it originates from the cross term $-V'U$
in the numerator of (\ref{force}) dominated by the product of
$\xi\varphi^2/2$ in $U$ and the derivative of the logarithmic
factor in $V'$. It is important that the resulting ratio in the
parenthesis of (\ref{force1}),
        \begin{eqnarray}
        \frac{\varphi^2}{\varphi_I^2}
        =\frac{\mbox{\boldmath$A$}/64\pi^2}{
        M_P^2/\xi\,\varphi^2}~,
        \end{eqnarray}
is in fact the ratio of two smallness parameters (\ref{smallness}).
This ratio is a priori not small even despite naively small quantum
corrections -- the result of multiplication of a small
$\mbox{\boldmath$A$}/32\pi^2$ by a very large $\xi$.

The meaning of restrictions (\ref{smallness}) becomes transparent in
the Einstein frame of fields $\hat g_{\mu\nu}$, $\hat\varphi$
related to the Jordan frame of (\ref{effaction}) by the equations
        \begin{eqnarray}
        &&\hat g_{\mu\nu}=\frac{2U(\varphi)}
        {M_P^2}g_{\mu\nu},                            \label{5.2}\\
        &&\left(\frac{d\hat\varphi}{d\varphi}\right)^2
        =\frac{M_P^2}{2}\frac{GU+3U'^2}{U^2}.      \label{5.3}
        \end{eqnarray}
The action (\ref{effaction}) in the Jordan frame, $\hat S[\hat
g_{\mu\nu},\hat\varphi]=S[g_{\mu\nu},\varphi]$, has a minimal
coupling, $\hat U=M_P^2/2$, canonically normalized inflaton field,
$\hat G=1$, and the new inflaton potential
        \begin{eqnarray}
        \hat{V}(\hat\varphi)=\left.\left(\frac{M_P^2}{2}\right)^2
        \frac{V(\varphi)}{U^2(\varphi)}
        \,\right|_{\,\varphi=\varphi(\hat\varphi)}~.     \label{hatV}
        \end{eqnarray}
This potential is very flat\footnote{Flatness of this potential
explains the cancellation of strongest classical terms in the
rolling force (\ref{force}) mentioned above. Note that the numerator
of (\ref{force}) is proportional to the gradient of the Einstein
frame potential (\ref{hatV}) which in the classical approximation
tends to a constant for $\varphi\to\infty$.} and monotonically
growing at least below the exponentially big value of the inflaton,
$\varphi<\varphi_*\simeq\mu^2\exp(192\pi^2\xi/\mbox{\boldmath$C$})$.
This provides us with a very big domain of the slow roll inflation.
The corresponding slow roll parameters for moderate (not
exponentially large) values of $\varphi$ read\footnote{Exponentially
big values of $\varphi$ near $\varphi_*$ and beyond can also
generate inflation which in particular becomes interminable at the
negative slope of $\hat V(\hat\varphi)$ for $\varphi>\varphi_*$. But
in this domain the one-loop approximation for radiative corrections
(\ref{effpot}) - (\ref{phirenorm}) breaks down definitely, not to
say about unobservable range of the relevant e-folding numbers.
Therefore we disregard this domain in what follows.}
    \begin{eqnarray}
    &&\hat\varepsilon \equiv\frac{M_P^2}2\left(\frac1{\hat
    V}\frac{d\hat V}{d\hat\varphi}\right)^2=
    \frac{4M_P^4}{3\xi^2\varphi^4}\,
    \left(1+\frac{\varphi^2}{\varphi_I^2}\right)^2=
    \frac43\left(
    \frac{M_P^2}{\xi\,\varphi^2}+
    \frac{\mbox{\boldmath$A$}}{64\pi^2}\!\right)^2,  \label{varepsilon}\\
    &&\hat\eta\equiv \frac{M_P^2}{\hat V}
    \frac{d^2\hat V}{d\hat\varphi^2}
    =-\frac{4M_P^2}{3\xi\varphi^2}~.      \label{eta}
    \end{eqnarray}
For $\varphi$ below $\varphi_I$ and for larger $\varphi$ the
smallness of these parameters reduces respectively to the first
and the second of restrictions (\ref{smallness}). Thus, these
restrictions are nothing but the conditions of sub-Planckian slow
roll inflation.

As it follows from (\ref{KG1}) and (\ref{force}) the e-folding
number of the inflation stage beginning with $\varphi$ and ending at
$\varphi_{\rm end}$ equals
    \begin{eqnarray}
    &&N = \int_{\varphi}^{\varphi_{\rm end}} d\varphi'\,
    \frac{3H^2(\varphi')}{F(\varphi')}=
    \frac{48\pi^2}{\mbox{\boldmath$A$}}
    \ln\frac{1+\varphi^2/\varphi_I^2}
    {1+\varphi_{\rm end}^2/\varphi_I^2}~.          \label{e-fold-q}
    \end{eqnarray}

This equation was used in \cite{BK} (Eq. (6.8)) with the initial
inflaton field $\varphi=\varphi_I$, the quantum scale of inflation
derived from the tunneling state of the Universe, and $\varphi_{\rm
end}\simeq 0$. When the both fields are small, $\varphi^2,
\varphi_{\rm end}^2\ll\varphi_I^2$, all the dependence on
$\mbox{\boldmath$A$}$ cancels out and $N$ reduces to Eq.(11) of
\cite{BS}
    \begin{eqnarray}
    N\simeq\frac{48\pi^2}{\mbox{\boldmath$A$}}
    \frac{\varphi^2-\varphi_{\rm end}^2}
    {\varphi_I^2}=\frac34\,\frac{\xi}{M_P^2}\,
    (\varphi^2-\varphi_{\rm end}^2)~.                               \label{NBezShap}
    \end{eqnarray}
However, only $\varphi_{\rm end}^2$ a priori satisfies this bound,
because it follows from $\hat\varepsilon\simeq 1$ it that
$\varphi_{\rm end}^2\simeq M_P^2/\xi$, and $\varphi_{\rm
end}^2/\varphi_I^2\simeq\mbox{\boldmath$A$}/64\pi^2\ll 1$
according to (\ref{smallness}). Therefore,
    \begin{eqnarray}
    &&N\simeq\frac{48\pi^2}{\mbox{\boldmath$A$}}
    \ln\left(1+\frac{\varphi^2}{\varphi_I^2}\right),
    \end{eqnarray}
and
    \begin{eqnarray}
    \frac{\varphi^2}{\varphi_I^2}=e^x-1,   \label{xversusvarphi}
    \end{eqnarray}
where we introduced a new parameter
    \begin{eqnarray}
    x\equiv\frac{N
    \mbox{\boldmath$A$}}{48\pi^2}.           \label{x}
    \end{eqnarray}
This parameter relates the initial value of the inflaton to the
quantum cosmological scale $\varphi^2_I$. In quantum cosmology the
normalizability of the cosmological quantum state requires
$\mbox{\boldmath$A$}$ to be positive \cite{qsi,BK}, so that
$\varphi_I^2>0$, and $\varphi=\varphi_I$ corresponds to $x=\ln 2$.
Here we adopt an alternative approach and deduce the value of $x$
from angular properties of CMB. In particular, we relax the
requirement of positivity for $\mbox{\boldmath$A$}$ and admit also
its negative values when the parameter $x$ is negative and
$\varphi^2<|\varphi_I^2|$ ($x\to - \infty$ for
$\varphi^2\to|\varphi_I^2|$).

In the limit of large $\varphi$ ($\varphi^2\gg \varphi_I^2>0$),
$N\propto \ln(|\varphi|/\varphi_I)$. Such dependence arises purely
due to one-loop quantum corrections to the potential $V$. Thus, in
this case one can indeed try to relate evolution of the Universe
during inflation to the renormalization group flow, in contrast to
usual inflation in the Einstein gravity where it is not possible
\cite{W08}.

\section{Primordial perturbation spectra and observational bounds}
Initial conditions for perturbations are chosen deep in the
WKB-regime before the first Hubble radius crossing during inflation,
where the perturbations themselves, as well as their energy density,
are conformally invariant approximately. On the other hand, at the
post-inflationary epoch the Einstein frame nearly coincides with the
Jordan one because the Higgs-inflaton settles at the minimum of the
classical potential $\varphi=\nu$ and $\nu^2\ll M_P^2/\xi$.
Therefore, observable cosmological perturbations can be directly
derived in the Einstein frame by standard formulae based on the slow
roll parameters (\ref{varepsilon})-(\ref{eta}) and the Einstein
frame potential (\ref{hatV}). In particular, the amplitude of
perturbations reads $\zeta^2(k)\equiv k^3\zeta_{{\bf k}}^2=\hat
V/24\pi^2 M_P^4\hat\varepsilon$, where the right-hand side is taken
at the moment $t=t(k)$ of the first Hubble radius crossing $k=aH$
that relates the comoving perturbation wavelength $k^{-1}$ to the
e-folding number $N$ from the end of inflation.

This expression can be also re-written in the form following from
the general $\delta N$ formalism \cite{AlStar1,S85,SS96} (see also
\cite{LR05,LSSTY05} for its recent developments):
\begin{equation}
\zeta^2=\left(\frac{dN}{d\varphi}\right)^2(\delta\varphi)^2~,~~
\delta\varphi=\frac{H}{2\pi}\frac{1}{\sqrt{1+\frac{3U'^2}{U}}}~.
\label{deltaN}
\end{equation}
Here $\delta\varphi\equiv (k^3\delta\varphi_{{\bf k}}^2)^{1/2}$ is
the rms fluctuation of a free minimally coupled scalar field in the
de Sitter background with the curvature $H$ -- the multiplier
$(1+3U'^2/U)^{-1/2}$ is due to a non-standard kinetic term in the
Einstein frame arising from the non-minimal coupling $U(\varphi)$ in
the physical (Jordan) frame, and we put $G=1$. \footnote{This result
is valid in both the Jordan and Einstein frames. The difference
between the number of e-folds in both frames is an effect of a
higher order in the slow-roll and loop expansions. In this
connection, see also a more detailed discussion of this topic in the
recent paper \cite{CY08} which appeared when the present paper was
prepared for submission.}

With (\ref{effpot}) and (\ref{varepsilon}) (or (\ref{e-fold-q}) as
well) in the range (\ref{smallness}), this gives the relation
    \begin{equation}
    \zeta^2=\frac{\lambda}{96\pi^2\xi^2\hat\varepsilon}=
    \frac{N^2}{72\pi^2}\,\frac\lambda{\xi^2}\,
    \left(\frac{e^x-1}{x\,e^x}\right)^2.
    \end{equation}
In view of the WMAP+BAO+SN normalization $\zeta^2\simeq 2.5\times
10^{-9}$ at the pivot point $k_0=0.002$ Mpc$^{-1}$ \cite{WMAPnorm}
which we choose to correspond to $N\simeq 60$, this yields the
following estimate on the ratio of coupling constants
    \begin{equation}
    \frac\lambda{\xi^2}\simeq
    0.5\times 10^{-9}\left(\frac{x\,e^x}{e^x-1}\right)^2.
    \end{equation}

Thus, the estimate $\lambda/\xi^2\sim 10^{-10}$ known from
\cite{Spokoiny,FM89,SalopekBondBardeen,FakirUnruh,qsi,BK,BS} is
modified here by the factor $(x\,e^x/(e^x-1))^2$ of entirely quantum
origin. This factor approaches unity for $x\ll 1$ but grows $\propto
x^2$ for $x\gg 1$. In quantum cosmology of the tunneling state
\cite{qsi,BK} with the initial value $\varphi=\varphi_I$, it is
equal to $(2\ln 2)^2\simeq 1.92$. As we will see now, observational
constraints lead to a much wider admissible range of $x$
corresponding to the observed window of scales.

The spectral index $n_s$ of the power spectrum of primordial
scalar (adiabatic) perturbations
    \begin{eqnarray}
    n_s&\equiv& 1+ \frac{d\ln \zeta^2(k)}{d\ln k}\approx 1 - \frac{d\ln
    \zeta^2(N)}{dN} \\ \nonumber
    &=& 1-
    \frac{2}{e^x-1}\,\frac{\mbox{\boldmath$A$}}{48\pi^2}=
    1-\frac{2}{N}\, \frac{x}{e^x-1}~. \label{ns}
    \end{eqnarray}
Note that $n_s=1-2/N$ for $x\ll 1 $, as in the $m^2\varphi^2$ or
$R+R^2/6M^2$ inflationary models (the latter model is the
simplified variant of the one introduced in \cite{S80}).

The power spectrum of primordial tensor perturbations
(gravitational waves) is
\begin{equation}
h_g^2(k)\equiv \!\!\!\!\sum_{\rm polarizations}\!\!\!\!k^3
<h_{\mu\nu}({\bf k})h^{\mu\nu}({\bf k})> = \frac{16G_{\rm
eff}H^2}{\pi}\approx \frac{V}{6\pi^2U^2}\approx
\frac{\lambda}{6\pi^2\xi^2}~, \label{hg}
\end{equation}
where $G_{\rm eff}$ is the effective large-scale Newton
gravitational constant in the physical frame. As a result, the
tensor-to-scalar ratio $r$ is given by the slow roll parameter
$\hat\varepsilon$ \cite{Mukhanov}:
    \begin{equation}
    r\equiv
    \frac{h_g^2}{\zeta^2}=16\hat\varepsilon=\frac{12}{N^2}\,
    \left(\frac{x e^x}{e^x-1}\right)^2~. \label{r}
    \end{equation}
Due to the $N^{-2}$ dependence, it is much smaller than in
$m^2\varphi^2/2$ and $\lambda\varphi^4/4$ models of inflation, but
it exactly coincides with the value of $r$ in the
$f(R)=M_P^2(R+R^2/6M^2)/2$ model \cite{S80,S83} for $x\ll 1$. The
fact that, in the limit $\varphi \ll |\varphi_I|$ when radiative
corrections are small, the model (\ref{effaction}) -
(\ref{phirenorm}) produces the same predictions for $n_s$ and $r$ as
this $f(R)$ model is a consequence of the latter model being
equivalent to the former one with $\lambda = \xi= M_P^2/3M^2$ (so
that the effective coupling constant $\lambda/\xi^2$ remains small)
and $G(\varphi)=\nu=\mbox{\boldmath$A$}=\mbox{\boldmath$B$}=
\mbox{\boldmath$C$}=\mbox{\boldmath$D$}=\mbox{\boldmath$E$}=
\mbox{\boldmath$F$}=0$. Note also that the consistency condition
$r=8n_t$ is satisfied in the model (\ref{effaction}-
\ref{phirenorm}), too, in spite of its non-Einsteinian (in fact,
scalar-tensor gravity) nature. To obtain it, the quantum one-loop
contribution to the potential $V$ should be taken into account.

Let us now compare the spectral index $n_s$ (\ref{ns}) to the
present observational data. Using the WMAP+BAO+SN constraint from
\cite{WMAPnorm} at the $2\sigma$ confidence level, we get
\begin{equation}
0.934 <n_s(k_0)<0.988~. \label{nsrange}
\end{equation}
Note that these bounds were obtained assuming $r=0$ (otherwise, they
have to be shifted up by about $0.01$). However, since in our model
$r$ appears to be much smaller than the upper $95\%$ confidence
level upper bound $r<0.2$ following from the same data, the estimate
(\ref{nsrange}) is just suitable for our purpose. For $N(k_0)=60$,
it leads to $0.36<x/(e^x-1)<1.98$ or $-1.57<x<1.79$. So,
$\mbox{\boldmath$A$}=48\pi^2x/N$ and $r$ belong to the following
ranges:
\begin{eqnarray}
-12.4<\mbox{\boldmath$A$}<14.1 ~,  \label{Arange} \\
0.0006<r<0.015~. \label{rrange}
\end{eqnarray}
Therefore, there is no "tensor desert" \cite{A07} in this model.

The corresponding spectral scalar index running
\begin{equation}
\alpha = - \frac{dn_s}{dN}=
-\frac{2}{N^2}\,\frac{x^2e^x}{(e^x-1)^2}=-\frac{e^x(n_s-1)^2}{2}
\label{running}
\end{equation}
is negative but negligible, and lies in the range $ -5.6 < \alpha
\times 10^4 < -4.3$ (the most negative value is reached for
$\mbox{\boldmath$A$}=0$). This is significantly lower than the
present observational upper bound on $|\alpha|$ (though the
negative sign of $\alpha$ is slightly favoured by data).

Thus, cosmic data leaves a rather wide window for possible values
of the quantity $\mbox{\boldmath$A$}$ and the tensor-to-scalar
ratio $r$ with the latter one varying by more than an order of
magnitude. This window includes also the case of inflation scale
generated by the tunneling state in quantum cosmology,
$\varphi=\varphi_I$, corresponding to $x\simeq 0.693$,
$\mbox{\boldmath$A$}=5.47$, $n_s=0.977$ and $r=0.0064$. This
window can be compared to constraints coming from the concrete
particle model -- the Standard Model with the Higgs field playing
the role of the inflaton.

\section{Standard Model bounds}
Since the particle masses induced by the Higgs effect scale as
$m(\varphi)\sim\varphi$, we can reduce the calculation of
$\mbox{\boldmath$A$}$, given by (\ref{Am}), to the present moment
when the Higgs field is in the vacuum state with $\varphi=\nu$.
Then the masses $m(\nu)$ comprise a well known set of the
presently observable $Z$ boson, $W^\pm$ boson and top quark
masses, $m_Z=91$ GeV, $m_W=80$ GeV and $m_t=171$ GeV, accompanied
by much lighter masses of other quarks and leptons giving a
negligible contribution. Taking into account three polarizations
for massive vector bosons and four polarizations times three
colors for quarks, we finally have
    \begin{eqnarray}
    \mbox{\boldmath$A$}=
    \frac2{\lambda \nu^4}{\rm tr}
    \sum_{\rm particles} (\pm 1)\, m^4(\nu)\simeq
    \frac6{\lambda \nu^4}\left(m_Z^4+2m_W^4-4m_t^4\right).
    \end{eqnarray}
The scale of the present symmetry breaking is known from the
measurement of the Fermi constant, $\nu=247$ GeV, while $\lambda$
can be expressed in terms of the Higgs mass, $m_H^2=2\lambda\nu^2$,
which is currently believed to be in the range 115 GeV$\leq m_H\leq$
180 GeV \cite{PDG}. Therefore, the total anomalous scaling constant
    \begin{eqnarray}
    \mbox{\boldmath$A$}\simeq
    \frac{12}{m_H^2 \nu^2}\left(m_Z^4+2m_W^4-4m_t^4\right)
    \end{eqnarray}
turns out to belong to the following range
    \begin{eqnarray}
    -48<\mbox{\boldmath$A$}<-20.     \label{SMArange}
    \end{eqnarray}

Unfortunately, it does not overlap with the range (\ref{Arange})
suggested by the CMB data. If the theory and future LHC
experiments on Standard Model could push the value of the Higgs
mass up to $\sim 230$ GeV, then SM generated inflation could yield
an observationally viable scenario with parameters at the lower
limit of the ranges in (\ref{nsrange}) and (\ref{rrange}), i.e.
with a low $n_s$ and a very low tensor-to-scalar ratio $r\simeq
0.0006$. \footnote{In this range of $m_H$ the inflaton
self-coupling becomes large, $\lambda\sim {\cal O}(1)$, so that
the running of coupling constants can slightly shift the above
numerical bounds -- the authors are grateful to F. Bezrukov for
this observation. The RG improved analysis of radiative
corrections in this range goes beyond the scope of this paper and
is currently under study.}

\section{Conclusions}
Thus, in principle the Standard Model Higgs can be the source of
inflation. However, the mechanism of this phenomenon is very
different from the suggestion of \cite{BS}, because it is dominated
by the quantum rather than by the tree-level part of the effective
action. In particular, the relation between the e-folding number and
the initial value of the inflaton (\ref{xversusvarphi}) is
determined by the parameter of the quantum anomalous scaling
$\mbox{\boldmath$A$}$. Therefore CMB data necessarily probe not only
the graviton-inflaton sector of SM or GUT type theory (\ref{model}),
but also all its heavy massive particles coupled to the inflaton.
The deviation of the CMB spectral index
    \begin{eqnarray}
    n_s(N)=1-\frac{2a}{e^{Na}-1}   \label{index}
    \end{eqnarray}
from unity is determined by the quantum conformal anomaly
$a\equiv\mbox{\boldmath$A$}/48\pi^2$.

Currently, however, this model of SM Higgs driven inflation seems
falsified. In particular, it strongly contradicts predictions of
quantum cosmology with the tunneling state, whose normalizability at
$\varphi\to\infty$ requires positive $\mbox{\boldmath$A$}$
\cite{qsi,BK,BKK}, though this conclusion might be revised in view
of the recent model of cosmological initial conditions in the form
of the microcanonical density matrix \cite{slih,why}.

More important is that the cosmological range (\ref{Arange}) of
$\mbox{\boldmath$A$}$ does not overlap with the SM range
(\ref{SMArange}) within the widely accepted rather strong upper
bound on the Higgs mass $m_H\leq 180$ GeV. This bound follows from
the arguments of \cite{PDG} justifying, in particular, the
electroweak perturbation theory with a small $\lambda\leq 0.26$.
Precision tests of this theory \cite{ALEPH} give at 95\%
confidence level a much weaker bound of 285 GeV which already
provides a big overlap with the cosmic range (\ref{Arange}),
$-12.4<\mbox{\boldmath$A$}<-8.0$. Within this overlap
    \begin{eqnarray}
    &&0.934<n_s<0.95,                    \label{combinednsrange}\\
    &&0.0006<r<0.001.                    \label{combinedrrange}
    \end{eqnarray}

Thus, it is up to the anticipated Higgs particle discovery at LHC,
which will or will not finally falsify the SM driven inflation. It
is important that in the latter case a possible tensor-to-scalar
ratio (\ref{combinedrrange}) is very small. This opens a big
reserve in future experiments for possible smallness of not yet
observed tensor perturbations.

Also, it should be noted that this inflationary model is a
representative of a broad class of models with a red tilted
spectrum and a small value of $r$ which belong neither to
"small-field", nor to "large-field" models, and are, in some
respects, intermediate between them. Namely, in the Einstein frame
these models have a practically constant potential in the
inflationary range, like small-field models, which however extends
over a large, or even semi-infinite, range of inflaton field
values, like in large-field models. The $R+R^2/6M^2$ model, as
well as the induced gravity model considered recently in
\cite{KSY08} fall into this class, too.

\section*{Acknowledgements}
The authors are grateful to F. Bezrukov and M. Shaposhnikov for
fruitful and thought-provoking correspondence and discussions. A.B.
is grateful for hospitality of the Perimeter Institute for
Theoretical Physics where a part of this work has been done. His
work was also supported by the RFBR grant 08-02-00725 and the grant
LSS-1615.2008.2. A.K. and A.S. were partially supported by the
RFBR grant 08-02-00923, the grant LSS-4899.2008.2 and by the Research
Programme "Elementary Particles" of the Russian Academy of Sciences.

\end{document}